\documentclass[%
 reprint,
superscriptaddress,
nofootinbib,
amsmath,amssymb,
aps,
prd,
]{revtex4-2}

\usepackage{graphicx}
\usepackage{dcolumn}
\usepackage{bm}

\usepackage{tikz}

\newcounter{comment}

\newcommand{\FScolor}{red}

{\refstepcounter{comment}%
\begin{quote}
\color{\PDcolor}{#1~ \normalsize \textbf{\underline{Comment} $\sharp$\thecomment~by~FP:~}}}%
{\end{quote}}

{\refstepcounter{comment}%
\begin{quote}
\color{\FScolor}{#1~ \normalsize \textbf{\underline{Comment} $\sharp$\thecomment~by~FS:~}}}%
{\end{quote}}

{\refstepcounter{comment}%
\begin{quote}
\color{\CMcolor}{#1~ \normalsize \textbf{\underline{Comment} $\sharp$\thecomment~by~Pp:~}}}%
{\end{quote}}

\newcommand{\fatg}{{\rm{I}}\!\Gamma}

\begin{document}


\title{A complete analysis of the Landau-gauge three-gluon vertex from lattice QCD}

\author{F.~Pinto-G\'omez}
\affiliation{Dpto. Ciencias Integradas, Centro de Estudios Avanzados en Fis., Mat. y Comp., Fac. Ciencias Experimentales, Universidad de Huelva, Huelva 21071, Spain}
\affiliation{Dpto. Sistemas F\'isicos, Qu\'imicos y Naturales, Univ. Pablo de Olavide, 41013 Sevilla, Spain}

\author{F.~De Soto}
\affiliation{Dpto. Sistemas F\'isicos, Qu\'imicos y Naturales, Univ. Pablo de Olavide, 41013 Sevilla, Spain}

\author{J.~Rodr\'iguez-Quintero}

\affiliation{Dpto. Ciencias Integradas, Centro de Estudios Avanzados en Fis., Mat. y Comp., Fac. Ciencias Experimentales, Universidad de Huelva, Huelva 21071, Spain}

\date{\today}

\begin{abstract}

Several continuum and lattice investigations of the QCD three-gluon vertex have recently exposed its key properties, some intimately connected with the low-momentum behavior of the two-point gluon Green's function and specially relevant for the emergence of a mass scale in this latter, \emph{via} the Schwinger mechanism. 
In the present study, we report on a lattice determination of the Landau-gauge, transversely projected three-gluon vertex, particularly scrutinizing an outstanding one of these properties, termed \emph{planar degeneracy}, exploring its implications and capitalizing on it to gain further insight on the low-momentum running of both the three-gluon vertex and its associated strong coupling.   

\end{abstract}

\maketitle

\section{Introduction}

The understanding of the origin of the mass is one of the most challenging, still open questions in particle physics. The Higgs mechanism
is well understood as the responsible for the masses of the electroweak interacting bosons and all leptons. However, concerning hadrons
and specially those composing stable matter, it can only explain just 1-2 $\%$ of the nucleons' masses. The hadrons are composite bound
states whose mass budget cannot be explained by the single contribution of their elementary components, the Lagrangian (current) quark
masses, and mostly relies on the dynamics of quantum chromodynamics (QCD), the Standard Model's strong interaction
theory\,\cite{Glashow:1961tr,Weinberg:1967tq,Salam:1968rm,Gross:1973id,Politzer:1973fx}. The hadron masses basically \emph{emerge} from
the interactions of quarks and gluons\,\cite{Roberts:2023lap}, although the nature of the interacting gluons is very much determined by the gluon-gluon interaction, which is an expression of a key feature of QCD: its non-abelian character. This same character triggers the anti-screening of color charges, which drives the interaction strength from its asymptotically free limit at low distances to a nonperturbative regime at low energies. And it is in this last that the rich and intriguing dynamics of QCD becomes apparent.       

The way in which gluon self-interactions define gluon properties is revealed by a central component of QCD, the tree-gluon vertex\,\cite{Marciano:1977su,Ball:1980ax,Davydychev:1996pb}. It has been established\,\cite{Binosi:2012sj,Aguilar:2011xe,Ibanez:2012zk,Binosi:2017rwj,Binosi:2022djx,Papavassiliou:2022wrb,Aguilar:2022thg} that the presence of longitudinally-coupled massless-pole structures in the three-gluon vertex activates the so-called Schwinger mechanism\,\cite{Schwinger:1962tn,Schwinger:1962tp,Cornwall:1981zr,Jackiw:1973tr,Eichten:1974et,Aguilar:2021uwa}, generating thereby a dynamical gluon mass and rendering the two-point gluon Green's function finite at vanishing momentum. This latter fact and that the ghost is transparent to the mass generation mechanism have been confirmed by a variety of studies of the fundamental QCD Green's functions carried out both in continuum\,\cite{Aguilar:2008xm,Boucaud:2008ky,Fischer:2008uz,RodriguezQuintero:2010wy,Pennington:2011xs,Maris:2003vk,Aguilar:2004sw,Fischer:2006ub,Kondo:2006ih,Boucaud:2006if,Binosi:2007pi,Binosi:2008qk,Boucaud:2007hy, 
Dudal:2007cw,Dudal:2008sp,Tissier:2010ts,Tissier:2011ey,Kondo:2011ab,Szczepaniak:2001rg,Szczepaniak:2003ve,Epple:2007ut,Szczepaniak:2010fe,Watson:2010cn,Watson:2011kv} and lattice QCD\,\cite{Cucchieri:2007md,Bogolubsky:2009dc,Oliveira:2009eh,Ayala:2012pb}.  

As the other side of the coin, when a dynamical gluon mass is generated, the interplay of massive gluon and massless ghost propagators, involved in the DSE expansions of QCD Green's functions, generates logarithmic singularities at the two- and three-point levels connected by the corresponding Slavnov-Taylor identities (STIs)\,\cite{Aguilar:2013vaa,Aguilar:2019jsj}. The two-point singularity comes from the ghost-loop contribution to the gluon self-energy, tamed by a kinematic factor $q^2$ and appearing only as a maximum in the gluon propagator form factor at non-zero momentum. However, it remains dynamically attached by the mass generation mechanism to the three-point singularity, associated to the tree-level tensor structure of the three-gluon vertex, where it is manifest in its corresponding form factor by crossing zero at low momentum, after an infrared suppression caused by the contributions from the usually called \emph{swordfish} diagrams\,\cite{Aguilar:2023qqd}. These properties for the three-gluon vertex have been unveiled by a series of recent works\,\cite{Cucchieri:2006tf,Cucchieri:2008qm,Pelaez:2013cpa,Aguilar:2013vaa,Boucaud:2013jwa,Blum:2014gna,Eichmann:2014xya,Mitter:2014wpa,Williams:2015cvx,Blum:2015lsa,Cyrol:2016tym,Athenodorou:2016oyh,Boucaud:2017obn,Aguilar:2021lke,Aguilar:2021okw,Duarte:2016ieu,Corell:2018yil,Aguilar:2019jsj,Aguilar:2019uob,Aguilar:2019kxz,Barrios:2022hzr,Pinto-Gomez:2022brg,Pinto-Gomez:2023zvj,Aguilar:2023qqd}, and have been revealed as instrumental for the formation of physically meaningful bound states\,\cite{Meyers:2012ka,Binosi:2014aea,Souza:2019ylx,Binosi:2016nme,Roberts:2020hiw,Huber:2018ned,Athenodorou:2020ani,Athenodorou:2021qvs,Huber:2021yfy}. 

Particularly in lattice QCD, the vertex form factors of two special classes of kinematic configurations, \emph{soft-gluon} and \emph{symmetric}, in which the vertex tensor space becomes reduced, have been extensively studied both with quenched \cite{Athenodorou:2016oyh,Boucaud:2017obn,Aguilar:2021lke,Aguilar:2021okw} and unquenched \cite{Boucaud:2013jwa} simulations in Landau gauge. Only very recently, a more general analysis extended to all available triplets of momenta with two equal squared ones, dubbed \emph{bisectoral} configurations, has been performed\,\cite{Pinto-Gomez:2022brg,Pinto-Gomez:2023zvj}. As a main outcome of this analysis, a three-gluon vertex property, termed therein \emph{planar degeneracy} and anticipated in some aspects by a previous DSE study\,\cite{Eichmann:2014xya}, is established. This property tells that, in very good approximation over a wide momentum range, the significant form factor is the one associated to the tree-level (classical) tensor and its behaviour is driven by only one Bose-symmetric combination of the three squared momenta. A first important consequence of this was soon exposed by a subsequent study\,\cite{Aguilar:2022thg}, in which a three-gluon vertex constructed with lattice QCD inputs by assuming planar degeneracy has been used to evaluate deviations in its corresponding STIs, shown to be consistent with the longitudinally-coupled massless poles triggering the Schwinger mechanism for the generation of a dynamical gluon mass.  

In the current work, for the first time, we extend the study of the three-gluon vertex obtained from lattice QCD in Landau gauge to a fully general kinematics, beyond the bisectoral class, furthermore implementing a complete,  improved tensor basis\,\cite{Aguilar:2023qqd}. We will initially focus on a proper analysis of lattice discretization artifacts and, next, scrutinize the deviations from planar degeneracy, paying special attention to the behavior of the non-classical form factors and their impact. Then, we will canvass the implications of this remarkable property and, capitalizing on it, elaborate further on the description of the non-perturbative running of the three-gluon vertex. An outstanding inference from planar degeneracy is its entailing a unique definition for the renormalized three-gluon vertex [\emph{viz.} Eq.\,\eqref{eq:3gluonR}] and effective coupling [\emph{viz.} Eq.\,\eqref{eq:alpha3g}], provided that the subtraction point is rationally defined. Furthermore, capitalizing on previous investigations of the soft-gluon and symmetric kinematic configurations from lattice QCD with 2+1 dynamical fermions simulated with a Domain-Wall action (DWF)\,\cite{Aguilar:2019uob}, we deliver the first unquenched results based on planar degeneneracy.

\section{The three-gluon vertex}

The starting step in our computational scheme is the calculation of the three-point correlation function in Fourier space $\langle\widetilde{A}_\alpha^a(q)\widetilde{A}_\mu^b(r)\widetilde{A}_\nu^c(p)\rangle$, where $\widetilde{A}_\alpha^a(q)$ corresponds to the gauge field Lorentz-color component \{$\alpha,a$\} at four-momentum $q$ and $\langle\cdots\rangle$ stands for the average over the Landau-gauge field configurations. More particularly, we will focus on its projection over the anti-symmetric color tensor $f^{abc}$,
\begin{equation}\label{eq:3gluonlat}
    \mathcal{G}_{\alpha\mu\nu}(q,r,p) = \frac{1}{24} f^{abc} \langle\widetilde{A}_\alpha^a(q)\widetilde{A}_\mu^b(r)\widetilde{A}_\nu^c(p)\rangle\,; 
\end{equation}
thereby discarding any possible contribution in the symmetric color tensor $d^{abc}$. 

\begin{figure}[!h]
\includegraphics[scale=0.35]{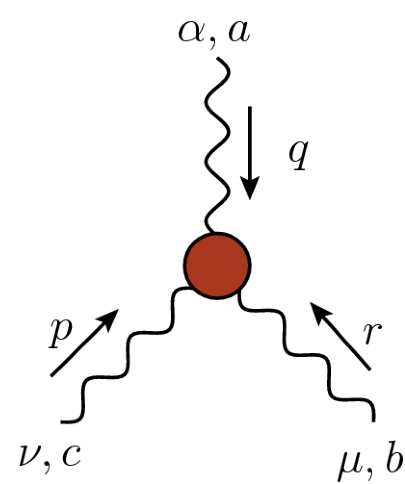} 
\caption{Diagramatic representation of the 1PI three-gluon vertex with the prescription chosen for the momenta and Lorentz-color indices. }
\label{fig:diagram} 
\end{figure}

The three gluon vertex depends on three momenta kinematically constrained by momentum conservation, $q+r+p=0$, what allows that any possible tensor should be formed by their different combinations, also including the metric tensor: \emph{e.g.}, $q_\alpha q_\mu q_\nu$ or $g_{\alpha\mu}p_\nu$. In total, there are 14 linearly independent tensors\,\cite{Ball:1980ax,Ball:1980ay}, although only 4 survive in Landau gauge owing to the transverse condition $q_\alpha \mathcal{G}_{\alpha\mu\nu}(q,r,p) = r_\mu \mathcal{G}_{\alpha\mu\nu}(q,r,p) = p_\nu \mathcal{G}_{\alpha\mu\nu}(q,r,p) = 0$.

As discussed at length in Ref.\,\cite{Athenodorou:2016oyh,Boucaud:2017obn,Aguilar:2021lke}, one can define the transverse projection of the one-particle irreducible (1PI) three-gluon vertex $\fatg^{\alpha' \mu' \nu'}$ (sketched in Fig.\,\ref{fig:diagram}), 
\begin{equation}\label{eq:Gammabardef}
\overline{\Gamma}_{\alpha\mu\nu}(q,r,p) = \fatg^{\alpha' \mu' \nu'}(q,r,p) P_{\alpha'\alpha}(q) P_{\mu'\mu}(r) P_{\nu'\nu}(p) \;, 
\end{equation}
with $P_{\mu\nu}(q)=g_{\mu\nu}-{q_\mu q_\nu}/{q^2}$ standing for the usual transverse projector; which can be thus derived from $\mathcal{G}_{\alpha\mu\nu}(q,r,p)$ as
\begin{equation}\label{eq:Gammabar}
    g\overline{\Gamma}_{\alpha \mu \nu}(q,r,p) = \frac{\mathcal{G}_{\alpha\mu\nu}(q,r,p)}{ \Delta(q^2) \Delta(r^2) \Delta(p^2) }
\end{equation}
where $g$ is the strong coupling and $\Delta(q^2)$ is defined from the gluon propagator as
\begin{equation}
    \langle \widetilde{A}^a_\mu(q) \widetilde{A}^b_\mu(-q) \rangle = \delta^{ab} P_{\mu\nu}(q) \Delta(q^2) \;.
\end{equation}
This \emph{transversely projected vertex} $\overline{\Gamma}_{\alpha\mu\nu}$ is the quantity that can be estimated from Landau-gauge lattice QCD simulations, from which the three-point correlation function in Eq.\,\eqref{eq:3gluonlat} can be calculated. It can be highlighted that it receives no contribution from the longitudinally coupled massless poles\,\cite{Aguilar:2011xe,Ibanez:2012zk,Binosi:2017rwj} shown to trigger the Schwinger mechanism\,\cite{Schwinger:1962tn,Schwinger:1962tp,Cornwall:1981zr,Jackiw:1973tr,Eichten:1974et,Aguilar:2021uwa,Aguilar:2022thg}. This transverse projection of the pole-free component of the 1PI three-gluon vertex is the object we will deal with in what follows.

\subsection{Bose-symmetric basis}
\label{subsec:Bose}

Embedded only within the 4-d transverse subspace of the tensor space defined by the 1PI three-gluon vertex, $\overline{\Gamma}_{\alpha\mu\nu}$ can be generally written in terms of any basis made by 4 linearly independent transverse tensors. Particularly, the choice of basis tensors expressing explicitly the Bose symmetry of the three-point Green function from Eq.\,\eqref{eq:3gluonlat} is advantageous, as has been detailed in Ref.\,\cite{Pinto-Gomez:2022brg} and will be again exploited below. A simple inspection of Eq.\,\eqref{eq:3gluonlat} makes apparent that, as $f^{abc}$ is anti-symmetric under exchange of color indices, so are too ${\cal G}_{\alpha \mu \nu}(q,r,p)$, and hence $\overline{\Gamma}_{\alpha \mu \nu}(q,r,p)$ defined by Eq.\,\eqref{eq:Gammabar}, under the simultaneous exchange of two momenta and their corresponding Lorentz indices. Then, one is left with\,\cite{Pinto-Gomez:2022brg}
\begin{equation}\label{eq:tr1PI}
    \overline{\Gamma}_{\alpha \mu \nu}(q,r,p) = \sum_{i=1}^4 \widetilde{\Gamma}_i(q^2,r^2,p^2) \widetilde{\lambda}_{i \alpha \mu \nu}(q,r,p) \,,
\end{equation}
with 
\begin{widetext}
\begin{subequations}
\label{eq:tls}
\begin{eqnarray}
\label{eq:tl1}
    \widetilde{\lambda}_{1\alpha \mu \nu}(q,r,p) &=& \left[ \ell_1^{\alpha'\mu'\nu'} + \ell_4^{\alpha'\mu'\nu'} + \ell_7^{\alpha'\mu'\nu'}\right] P^{\alpha'}_\alpha(q)  P^{\mu'}_{\mu}(r)  P^{\nu'}_\nu(p) \,,\\
\label{eq:tl2}
    \widetilde{\lambda}_{2\alpha \mu \nu}(q,r,p) &=& 3 \frac{(q-r)_{\nu'} (r-p)_{\alpha'} (p-q)_{\mu'}}{q^2+r^2+p^2}  
    P^{\alpha'}_\alpha(q)  P^{\mu'}_{\mu}(r)  P^{\nu'}_\nu(p)\,,\\
\label{eq:tl3}
\widetilde{\lambda}_{3\alpha \mu \nu}(q,r,p) &=& \frac{3}{q^2+r^2+p^2}   
\left[\ell_3^{\alpha'\mu'\nu'} + \ell_6^{\alpha'\mu'\nu'} + \ell_9^{\alpha'\mu'\nu'}\right] P_{\alpha'}^\alpha(q)  P_{\mu'}^{\mu}(r)  P_{\nu'}^\nu(p) \,,\\
\label{eq:tl4}
\widetilde{\lambda}_{4\alpha \mu \nu}(q,r,p) &=& \left( \frac{3}{q^2+r^2+p^2}\right)^2
\left[t_1^{\alpha\mu\nu} + t_2^{\alpha\mu\nu} + t_3^{\alpha\mu\nu}\right]\,,
\end{eqnarray}
\end{subequations}
\end{widetext}
where $t_i$ and $l_j$ stand, respectively, for the 4 transverse and 10 non-transverse, well-known Ball-Chiu tensors\,\cite{Ball:1980ax,Ball:1980ay} (see, \emph{e.g.}, Eqs.\,(3.4) and (3.6) of Ref.\,\cite{Aguilar:2019jsj}). The basis tensors given by Eqs.\,\eqref{eq:tls} can be proven to express the anti-symmetric character of $\overline{\Gamma}_{\alpha \mu \nu}(q,r,p)$ under Bose transforations: \emph{e.g.}, $\widetilde{\lambda}_{i\alpha \mu \nu}(q,r,p) = - \widetilde{\lambda}_{i \mu \alpha \nu}(r,q,p)$. However, keeping this anti-symmetric property, other choices are possible and, in some aspects, preferable. 

Let us generally consider another tensor basis such that, 
\begin{align}
\label{eq:tr1PIast}
    \overline{\Gamma}_{\alpha \mu \nu}(q,r,p) = \sum_{i=1}^4 \widetilde{\Gamma}^\ast_i(q^2,r^2,p^2) \widetilde{\lambda}^\ast_{i \alpha \mu \nu}(q,r,p) \,  
\end{align}
where one and another form factors can be related as
\begin{subequations}\label{eq:Proj0}
\begin{align} 
\widetilde{\Gamma}^\ast_i(q^2,r^2,p^2) &= {\cal P}_i^{\ast \alpha\mu\nu} \, \overline{\Gamma}_{\alpha \mu \nu}(q^2,r^2,p^2)  \label{eq:Proj} \\
&= \sum_{k=1}^4 {\cal P}_i^{\ast \alpha\mu\nu} \, \widetilde{\lambda}_{k \alpha \mu \nu}(q,r,p) \, \widetilde{\Gamma}_k(q^2,r^2,p^2) \label{eq:transf}
\end{align}    
\end{subequations}
where Eq.\,\eqref{eq:Proj} displays the form factors projected out from $\overline{\Gamma}_{\alpha \mu \nu}(q,r,p)$, as it is expanded in \eqref{eq:tr1PIast}, with the projector
\begin{align}\label{eq:Projdef}
{\cal P}_i^{\ast \alpha\mu\nu} = \sum_{j}^4 \widetilde{M}^{\ast -1}_{ij}(q^2,r^2,p^2) \widetilde{\lambda}^{\ast \alpha \mu \nu}_j(q,r,p) \;,     
\end{align}
defined by the $4\times 4$ matrix 
\begin{align}\label{eq:Mm1}
\widetilde{M}_{ij}^\ast (q^2,r^2,p^2) = 
\widetilde{\lambda}^{\ast \alpha \mu \nu}_i(q,r,p) \widetilde{\lambda}^\ast_{j \alpha \mu \nu}(q,r,p) \;;  
\end{align}
while \eqref{eq:transf} results from replacing the vertex with its expansion given by Eq.\,\eqref{eq:tr1PI}. 

The basis defined by Eqs.\,\eqref{eq:tls} has been introduced in Ref.\,\cite{Pinto-Gomez:2022brg} as a generalisation of the restricted one employed in Refs.\,\cite{Athenodorou:2016oyh,Boucaud:2017obn,Aguilar:2021lke}. Indeed, $\widetilde{\lambda}_{1 \alpha \mu \nu}(q,r,p)$ corresponds to the three-gluon tree-level tensor which, together with $\widetilde{\lambda}_{2 \alpha \mu \nu}(q,r,p)$, form a basis for those special kinematic configurations in which the three squared momenta are equal (\emph{symmetric} case); while $\widetilde{\lambda}_{3 \alpha \mu \nu}(q,r,p)$ and $\widetilde{\lambda}_{4 \alpha \mu \nu}(q,r,p)$ were therein chosen for the basis completion in the 4-d transverse subspace. An alternative choice\,\cite{Aguilar:2023qqd} can be featured as
\begin{align}\label{eq:redef}
\widetilde{\lambda}^\ast_{i \alpha \mu \nu}(q,r,p) =  \widetilde{\lambda}_{i \alpha \mu \nu}(q,r,p) - \delta_{i3} \frac 3 2 \widetilde{\lambda}_{1 \alpha \mu \nu}(q,r,p) \;;  
\end{align}
implying that Eq.\,\eqref{eq:transf} specializes as 
\begin{align}\label{eq:GastvsG}
\widetilde{\Gamma}^\ast_i(q^2,r^2,p^2) = \sum_{k=1}^4 \left( \delta_{ik} + \frac 3 2 \delta_{i1}\delta_{k3} \right) \widetilde{\Gamma}_k(q^2,r^2,p^2) \;.  
\end{align}
The latter will be shown below as a very useful result entailing that, after the redefinition of the basis by Eq.\,\eqref{eq:redef}, the form factor $\widetilde{\Gamma}^\ast_1(q^2,q^2,0)$ fully delivers the \emph{soft-gluon} limit ($p^2 \to 0$) of the transversely projected vertex\,\cite{Aguilar:2023qqd}.    

\subsection{Kinematics}

Owing to momentum conservation, a kinematic configuration for the three-gluon vertex remains entirely determined by the three squared momenta or, alternatively, by two momenta and the angle they form (\emph{e.g.}, $q^2,r^2$ and $\theta_{qr}=[p^2-q^2-r^2]/[2\sqrt{q^2 r^2}]$; or any completely analogous combination). Furthermore, the scalar form-factors defined in Eqs.\,(\ref{eq:tr1PI},\ref{eq:tr1PIast}), which depend on three scalars, can be equivalently characterized. Interestingly, the basis being Bose-symmetric, they can only depend on Bose invariants that can be recast as\footnote{For reference it is convenient to write these invariants in terms of the ones defined in Eqs.(50) and (53) of \cite{Eichmann:2014xya}, $\mathcal{S}_0$, $\mathcal{S}_1$, and $\mathcal{S}_2$: $s^2=3\mathcal{S}_0$, $t^4=6\mathcal{S}_0^2\mathcal{S}_1$ and $u^6=54\mathcal{S}_0^3\mathcal{S}_2$.}
\begin{subequations}\label{eq:Bose-inv}
\begin{align}
    s^2 &= \frac{q^2+r^2+p^2}{2} \;, \label{eq:s2}\\
    t^4 &= \frac{(q^2-r^2)^2+(r^2-p^2)^2+(p^2-q^2)^2}{3} \;, \label{eq:t4} \\
    u^6 &= (q^2+r^2-2p^2)(r^2+p^2-2q^2)(p^2+q^2-2r^2) \;. \label{eq:u6}
\end{align}    
\end{subequations}
An extensive analysis of the three-gluon kinematic configurations is done in Ref.\,\cite{Eichmann:2014xya} on the ground of the theory of permutation group; while a geometric re-derivation can be found in Ref.\,\cite{Pinto-Gomez:2022brg}. In the aim of keeping enough self-completion in the present document, we will shortly outline their main outcomes.  

\begin{figure}[!t]
\begin{tabular}{cc}
\begin{tabular}{c}
\rule[0cm]{0cm}{1cm} \\
		\begin{tikzpicture}[scale=0.8]
			\node (r) at ( 3.0,  0.0) {}; %
			\node (q) at (-3.0,  0.0) {}; %
			\node (p) at ( 0.0, 5.36) {}; %
			\node (p2) at ( 0.0, 3.55) {}; 
			\node (r2) at ( -1.57, .832) {}; 
			\node (q2) at (  1.57, .832) {}; 
			\node (S) at (0,1.75) {};
			\node (Op) at (0,2.625) {};
			\node (Oq) at (-0.76,1.31) {};
			\node (Or) at (0.76,1.31) {};
			\node (SGp) at (0,0) {};
			\node (SGq) at (1.52,2.63) {};
			\node (SGr) at (-1.52,2.63) {};
			\node (apA) at ( 3.0,  1.75) {}; %
			\node (amA) at ( -3.0,  1.75) {}; %
			\node (apB) at ( 0,  6.25) {};
			\fill[fill=blue!20] (p.center) -- (r.center) -- (q.center) -- (p.center);	-
			\draw [thick, blue] (r) -- (q) -- (p) -- (r);
			\fill[fill=white] (S) circle (1.72cm);
			\draw [thick,dashed,pink] (S) -- (SGr);
			\draw [thick,dashed,pink] (S) -- (SGq);
			\draw [thick,dashed,pink] (S) -- (SGp);
			\draw [very thick,gray] (p2) -- (S);
			\draw [very thick,gray] (r2) -- (S);
			\draw [very thick,gray] (q2) -- (S);
			\node at (-0.4,2.05) {$S$};
			\node at (S)[circle,fill,inner sep=1.5pt,green]{};
			\node at (0,-0.4) {${p^2=0}$};
			\node at (SGp) [circle,fill,inner sep=1.5pt,orange]{};
			\node at (2.5,2.8) {${q^2=0}$};
			\node at (SGq) [circle,fill,inner sep=1.5pt,black]{};
			\node at (-2.5,2.8) {${r^2=0}$};
			\node at (SGr) [circle,fill,inner sep=1.5pt,black]{};
		\end{tikzpicture} 
\end{tabular}
&
\hspace*{-1.5cm}
\begin{tabular}{c}
		\begin{tikzpicture}[scale=0.4]
			\fill[fill=blue!30] (0,3) -- (3,0) -- (-1.25,-1.25) -- (0,3);
			\draw [very thick, -latex] (0,0) -- (0,5) node [right] {$p^2$};
			\draw [very thick, -latex] (0,0) -- (-2.5,-2.5) node [right] {$q^2$};
			\draw [very thick, -latex] (0,0) -- (5,0) node [right] {$r^2$};
			\draw [thick, blue] (0,3) -- (3,0) -- (-1.25,-1.25) -- (0,3);
		\end{tikzpicture}
		\\
		\rule[0cm]{0cm}{3.5cm}		
\end{tabular}		 
\end{tabular}
\caption{The kinematic configurations for the three-gluon vertex sketched in Fig.\,\ref{fig:diagram}, represented here by the the Cartesian coordinates $(q^2,r^2,p^2)$. A fixed vaue of $s^2$, Eq.\,\eqref{eq:s2}, implies sitting on the same plane perpendicular to the octant diagonal (top-right), which defines an equilateral triangle in which momentum conservation restricts the representations to lie within the white incircle (left). The green dot labels the symmetric case, while black and orange stand for the soft-gluon, the latter specializing the case considered, and connected by permutations with the other two. The explanation of this representation is expanded in the text.       
}
\label{fig:triangle}
\end{figure}
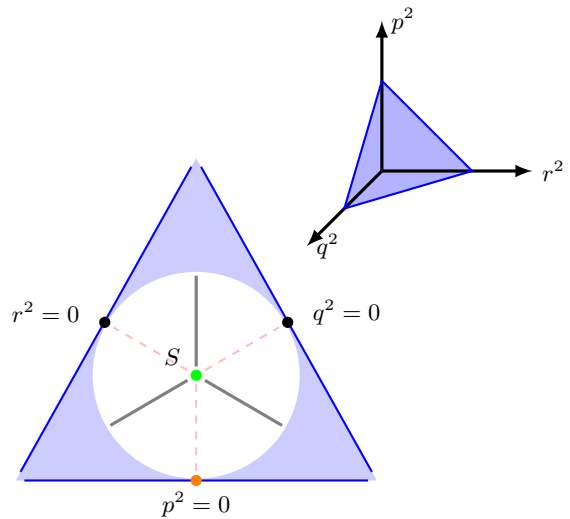

Any possible kinematical configuration, characterized by the three squared momenta, can be represented as a point in the positive octant of a 3-d space, whose Cartesian axes correspond to $q^2$, $r^2$ and $p^2$ (see the top-right plot of Fig.~\ref{fig:triangle}). All configurations sharing the same $s$-invariant, Eq.\,\eqref{eq:s2}, sit on a plane perpendicular to the diagonal of this positive octant (the distance of the plane to the origin along the diagonal is $2s^2/\sqrt{3}$), which defines an equilateral triangle (drawn in blue in the figure). Furthermore, momentum conservation, $q+r+p=0$, imposes a restriction on the momenta such that all allowed configurations lie within the triangle incircle, the center of which corresponds to the symmetric configuration: $q^2=r^2=p^2$ (represented by a green dot). The $t$-invariant, Eq.\,\eqref{eq:t4} locates the point representing the configuration on a given concentric circumference within the incircle, and expresses its separation away from the symmetric case (momentum conservation translates into $t^2 \le \sqrt{2/3} s^2$). And, finally, the $u$-invariant, Eq.\,\eqref{eq:u6}, refers the location to each of the three bisectoral lines in the triangle. As discussed in Ref.\,\cite{Pinto-Gomez:2022brg}, all the configurations such that two squared momenta and two angles are the same can be represented by a piece of bisectoral line lying inside the incircle (\emph{e.g.}, the vertical one in Fig.\,\ref{fig:triangle} corresponds to $r^2=q^2$ and $\theta_{qp}=\theta_{rp}$). Along any of the three bisectoral segments, the kinematics evolves from soft-gluon to collinear non-soft cases, with the symmetric one in between; \emph{e.g.}, from the cases $p=0$ and $\theta_{qp}=\theta_{rp}=\pi/2$ (orange point) to $q^2=r^2=p^2/4$ and $\theta_{qp}=\theta_{rp}=\pi$ (in Landau gauge, no tensor structure remains for this case), passing through $q^2=p^2=r^2$ and $\theta_{pq}=\theta_{rp}=2\pi/3$ (green point), as illustrated in the vertical bisectoral of Fig.\,\ref{fig:triangle}. And, accordingly, the $u$-invariant evolves from $u^6=2s^6$ (soft-gluon) to $u^6=-2s^6$ (collinear non-soft), becoming zero in the symmetric case. 

It is worthwhile to underline that the incircle representing all allowed configurations can be divided into three identical regions, covering each an angle $2\pi/3$, that can be mapped into each other by permutations of the squared momenta $q^2$, $r^2$ and $p^2$, which do not modify the Bose-symmetric invariants defined in Eq.\,\eqref{eq:Bose-inv}. In Fig.\,\ref{fig:triangle}, these three regions appear bordered by solid grey lines displaying the bisectoral segments which join the symmetric and collinear non-soft cases. On top of this, Bose symmetry\footnote{the operation of reflection in respect to the bisectoral segment joining the symmetric and the $p=0$ case correspond to the exchange $q^2 \leftrightarrow r^2$; completely analogous properties hold for the other two bisectoral segments.} also guarantees that the invariants remain unmodified by the reflection in respect to the bisectoral segment joining symmetric and soft-gluon cases, drawn with a pink dashed line in the figure. Therefore, only the kinematic configurations represented by points within one of the six regions delimited by solid grey and dashed pink lines, spanned by an angle $\pi/3$, are independent.

In general, the transversely projected vertex can have independent contributions in the four basis tensors defined by either Eqs.~\eqref{eq:tls} or \eqref{eq:redef}, with the corresponding form factors projected out by applying Eq.\,\eqref{eq:Proj} and, in the former case, also Eq.\,\eqref{eq:GastvsG}. They may depend on the three invariants \eqref{eq:Bose-inv} that can be evaluated for any arbitrary kinematic configuration. 

However, we will begin by specializing for the bisectoral kinematics, as in Ref.\,\cite{Pinto-Gomez:2022brg}, highlighting the particular soft-gluon and symmetric cases. With no loss of generality, one can then focus on the case $q^2=r^2$ and $\theta_{pq}=\theta_{pr}$, for any $p^2$. The sub-space spanned by the transversely projected vertex reduces thereby its dimension from 4 down to 3, and the new restricted tensor basis obtained from Eq.\,\eqref{eq:redef} is formed by the three tensors 
\begin{align}
\overline{\lambda}^\ast_{i \alpha\mu\nu}(q,r,p) = \lim_{r^2 \to q^2}\widetilde{\lambda}^\ast_{i \alpha\mu\nu}(q,r,p) \;,   
\end{align}
for $i=1,2,3$; while 
\begin{align}
\lim_{r^2\to p^2} \widetilde{\lambda}^\ast_{4 \alpha\mu\nu}(q,r,p) = 
\sum_{i=1}^3 f^\ast_i\left(\frac{2p^2}{p^2+2q^2}\right) \widetilde{\lambda}^\ast_{i \alpha\mu\nu}(q,r,p) \;,
\end{align}
with
\begin{align}
f_1^\ast(z)= \frac{9}{16} z (2-z) ,\; f_2^\ast(z)= \frac 3 8 \left(\frac 3 4 z - 1\right) ,\; f_3^\ast(z) = \frac 3 8 z \;.
\end{align}
A restricted basis of tensors $\overline{\lambda}_{i \alpha\mu\nu}$ can be analogously derived from \eqref{eq:tls}, and formally identical results obtained with the only difference that $f_1(z)=9/16 z(1-z)$\,\cite{Pinto-Gomez:2022brg}.   

Then, Eqs.\,(\ref{eq:Proj0}-\ref{eq:Mm1}) can be specialized to the bisectoral case only with the replacement of 4-d by 3-d basis tensors, restricting the running of $i$, $j$ and $k$ from 1 to 3. This entails for the non-invertible $4\times 4$ matrix \eqref{eq:Mm1} its being replaced by its first $3 \times 3$ block in the limit $r^2 \to q^2$, 
\begin{align}\label{eq:Mbar}
\overline{M}^\ast_{ij}(q^2,p^2) = \lim_{r^2\to q^2} \widetilde{M}^\ast_{ij}(q^2,r^2,p^2)    
\end{align}
for $i,j=1,2,3$; the form factors for the restricted basis reading now 
\begin{align}\label{eq:barfromtilde}
\overline{\Gamma}^\ast_i(q^2,p^2) =& \lim_{r^2\to q^2} 
\widetilde{\Gamma}^\ast_i(q^2,r^2,p^2) \nonumber \\ 
&+ f_i^\ast\left(\frac{2p^2}{p^2+2q^2}\right) \widetilde{\Gamma} ^\ast_4(q^2,r^2,p^2) \;,
\end{align}
in terms of those for the general basis. And a formally identical equation holds relating the form factors $\widetilde{\Gamma}_i$'s and $\overline{\Gamma}_i$'s defined from Eqs.\,(\ref{eq:tr1PI},\ref{eq:tls}). Analogously, in the bisectoral case, an equation equivalent to \eqref{eq:GastvsG} works to make $\overline{\Gamma}^\ast_i$'s read in terms of $\overline{\Gamma}_k$'s, with $i,k=1,2,3$.       

The symmetric ($p^2=q^2=r^2$, $\theta_{pq}=\theta_{pr}=2\pi/3$) and soft-gluon ($p^2=0$, $\theta_{pq}=\theta_{pr}=\pi/2$) are special cases in which the $3\times 3$ matrix defined in Eq.\,\eqref{eq:Mbar} is non-invertible, as the involved tensor sub-spaces take dimensions 2 and 1, respectively, making its determinant vanish. 

In the symmetric case, the new restricted basis can be defined as [$i=1,2$]
\begin{align}
\lambda_{i \alpha\mu\nu}(q,r,p) = \lim_{p^2 \to q^2} \overline{\lambda}_{i \alpha\mu\nu}(q,r,p) = \lim_{p^2 \to q^2} \overline{\lambda}^\ast_{i \alpha\mu\nu}(q,r,p)   \;;
\end{align}
in terms of which, the transversely projected vertex reads
\begin{align}
\overline{\Gamma}_{\alpha\mu\nu}(r,q,p)= \sum_{i=1}^2 \overline{\Gamma}^\textrm{sym}_i (q^2) \lambda_{i \alpha\mu\nu}(q,r,p) \;;    
\end{align}
with
\begin{subequations}
\label{eq:Gamma12sym}
\begin{align}
    \overline{\Gamma}_1^\text{sym}(q^2) = &\lim_{p^2\to q^2} \overline{\Gamma}_1(q^2,p^2) + \frac 1 2  \overline{\Gamma}_3(q^2,p^2) \\ 
    & \lim_{p^2\to q^2} \overline{\Gamma}^\ast_1(q^2,p^2) -  \overline{\Gamma}^\ast_3(q^2,p^2) \;, \\
    \overline{\Gamma}_2^\text{sym}(q^2) = &\lim_{p^2 \to q^2} \overline{\Gamma}_2(q^2,p^2) - {\frac{3}{4}} \overline{\Gamma}_3(q^2,p^2)  \\
    = &\lim_{p^2 \to q^2} \overline{\Gamma}^\ast_2(q^2,p^2) - {\frac{3}{4}} \overline{\Gamma}^\ast_3(q^2,p^2) \;.
\end{align}    
\end{subequations}

On the other hand, in the soft-gluon case, one is left only with the tree-level tensor $\lambda_{1 \alpha\mu\nu}(q,r,p)$, which corresponds to $\overline{\lambda}^\ast_{1 \alpha\mu\nu}(r,q,p)$=$\overline{\lambda}_{1 \alpha\mu\nu}(r,q,p)$ taken in the limit $p\to 0$; and can thus write
\begin{align}\label{eq:Gammasg}
\overline{\Gamma}_{\alpha\mu\nu}(r,q,p)= \overline{\Gamma}^\textrm{sg} (q^2) \lambda_{1 \alpha\mu\nu}(q,r,p) \;;    
\end{align}
with
\begin{subequations}\label{eq:sg}
\begin{align}\label{eq:softlimit}
    \overline{\Gamma}^\text{sg}(q^2) &= \lim_{p^2\to 0} \overline{\Gamma}_1(q^2,p^2) + \frac 3 2 \overline{\Gamma}_3(q^2,p^2) \\ 
    \label{eq:softlimitast}
    &= \lim_{p^2\to 0} \overline{\Gamma}^\ast_1(q^2,p^2) \;. 
\end{align}    
\end{subequations}
The outcome exhibited by Eq.\,\eqref{eq:sg}, which can be traced back to Eq.\,\eqref{eq:GastvsG}, appears as the main advantage of the tensor basis \eqref{eq:redef}, Ref.\,\cite{Aguilar:2023qqd}, in respect to \eqref{eq:tls}, Ref.\,\cite{Pinto-Gomez:2022brg}. Namely, the unique form factor in the soft-gluon case is only given by the appropriate limit of the tree-level tensor form factor; while, is the basis \eqref{eq:tls} used, two form factors will need to be combined to deliver the correct result. As it is raised in Ref.\,\cite{Pinto-Gomez:2022brg}, both $\overline{\Gamma}_1(q^2,p^2)$ and $\overline{\Gamma}_3(q^2,p^2)$ are separately plagued by a perturbative divergence in the limit $p\to 0$, which however cancels out when they are combined as in Eq.\,\eqref{eq:softlimit}, rendering finite and well-defined the soft-gluon form factor and, owing to \eqref{eq:softlimitast}, the bisectoral $\overline{\Gamma}^\ast_1(q^2,p^2)$. These divergences in $\overline{\Gamma}_{1,2}(q^2,p^2)$ have been seen to be individually tamed by the nonperturbative generation of a gluon mass\,\cite{Aguilar:2022thg,Aguilar:2023qqd}. However, as will be discussed below, their remnant seems to spoil in some kinematic region some of the good properties of the three-gluon form factors, making specially useful their being removed by the implementation of the tenson basis \eqref{eq:redef}.

\subsection{Renormalization of the three-gluon form-factors and coupling.}

We have thus far elaborated on the derivation of the relevant form factors for the transversely projected vertex $\overline{\Gamma}_{\alpha\mu\nu}$ from the three-point Green's function ${\cal G}_{\alpha\mu\nu}$. The latter is herein computed from lattice QCD and, thereby, obtained as a lattice-regularized bare quantity. A multiplicative renormalization scheme is subsequently applied 
by introducing first the gauge field renormalization constant $Z_A$, such that bare and renormalized gluon two- and three-point Green functions relate as
\begin{subequations}\label{eq:ZAdef}
\begin{align}\label{eq:ZAdef2p}
  \Delta_R(q^2) &=\lim_{a\to 0} Z_A^{-1}(a) \Delta(q^2;a),  \\
  {\cal G}_{R\alpha\mu\nu}(q,r,p) &= \lim_{a\to 0}Z_A^{-3/2}(a) {\cal G}_{\alpha\mu\nu}(q,r,p;a); 
  \label{eq:ZAdef3p}
\end{align}    
\end{subequations}
where we have made explicit the dependence on the regularization scale, the lattice spacing $a$, of the bare quantities and renormalization constants; and  have left implicit the latter's and renormalized quantities' on the subtraction scale $\zeta^2$. Capitalizing on multiplicative renormalizability, only one further renormalization constant needs to be introduced for all the form factors of the 1PI three-gluon vertex, 
\begin{align}\label{eq:Z3def}
  \widetilde{\Gamma}_{i R}^\ast(q^2,r^2,p^2) &= \lim_{a\to 0} Z_3(a) \widetilde{\Gamma}_i^\ast(q^2,r^2,p^2;a)\,;
\end{align}
thus implying
\begin{align}\label{eq:Zgdef}
g_R = \lim_{a\to 0} Z_A^{3/2}(a) Z_3^{-1}(a) \, g(a)      
\end{align}
for the renormalization of the strong coupling. The same reads for the form factors from Eq.\,\eqref{eq:tr1PI} for the tensor basis \eqref{eq:tls}.  

The last step of the calculation program is the implementation of a renormalization prescription; namely, \emph{momentum subtraction} (MOM)\,\cite{Hasenfratz:1980kn} in our case. This implies that all renormalized correlation functions acquire their tree-level expressions at the subtraction point, defined in terms of the subtraction scale, $\zeta^2$. In particular, one sets \mbox{$\Delta^{-1}_{{\rm R}}(\zeta^2) = \zeta^2$}, which fixes $Z_A$, 
\begin{align}\label{eq:ZAsol}
Z_A(a^2) = \zeta^2 \Delta(\zeta^2;a) \;;    
\end{align}
while, for the the three-gluon Green's function, one needs to consider a given kinematic configuration, $Z_3$ resulting defined by 
\begin{align}
\left. \widetilde{\Gamma}_{1 R{\textrm{k}(\zeta)}}^\ast(q^2,r^2,p^2) \right|_{\textrm{k}(\zeta)} &= 
\lim_{a\to 0}  Z_3^{\textrm{k}(\zeta)}(a)  
\left. \widetilde{\Gamma}_1^\ast(q^2,r^2,p^2;a) \right|_{\textrm{k}(\zeta)} 
\nonumber \\ \label{eq:Z3kc}
&= 1 \;,
\end{align}
where ${\textrm{k}(\zeta)}  \equiv \{\zeta^2,\theta_{qp},\theta_{rp}\}$ specifies the chosen configuration and $R{\textrm{k}(\zeta)}$ its associated renormalization scheme. At this point, we take advantage of the three-gluon kinematic analysis described in the previous subsection and choose a configuration represented by a point lying on the plane $s^2=\zeta^2$, within the white circle of Fig.~\ref{fig:triangle}, fixed by two angles $\theta_{qp}$ and $\theta_{rp}$.   

In Eqs.\,(\ref{eq:ZAdef}-\ref{eq:Zgdef},\ref{eq:Z3kc}), the limit $a\to 0$, required to drop any subleading lattice artifact away, has been made explicit. This is a mandatory part of a renormalization program, intended to remove properly any remaining non-singular dependence on the regularization scale or cut-off, after subtraction. In lattice QCD calculations, the latter entails an extrapolation to the continuum limit which, for the gluon and ghost propagators can be done by following the procedure detailed in \cite{Boucaud:2018xup}. However, when three-points functions are involved, The subleading lattice artifacts, once those related to $O(4)$-breaking are properly treated (as will be discussed below), become hidden by the statistical errors and, even at non-zero but small lattice spacing, the renormalized quantities appear not to depend on it. The limit $a \to 0$ can be then removed in practice but, formally at least, Eq.\,\eqref{eq:ZAsol} and  
\begin{align}\label{eq:Z3kcdef}
Z_3^{\textrm{k}(\zeta)}(a) = \left(\left. \widetilde{\Gamma}_1^\ast(q^2,r^2,p^2;a)\right|_{\textrm{k}(\zeta)} \right)^{-1}     
\end{align}
are only possible solutions, all differing only by subleading $\mathcal{O}(a^2)$ artifacts, of \eqref{eq:ZAdef2p} evaluated at $q^2=\zeta^2$ and Eq.\,\eqref{eq:Z3kc}, respectively. They imply 
\begin{align}\label{eq:GiR}
\widetilde{\Gamma}_{iR{\textrm{k}(\zeta)}}^\ast(q^2,r^2,p^2) = \lim_{a \to 0} 
\frac{\widetilde{\Gamma}_i^\ast(q^2,r^2,p^2;a)}
{\left. \widetilde{\Gamma}_1^\ast(q^2,r^2,p^2;a) \right|_{\textrm{k}(\zeta)}}\;, 
\end{align}
for $i=1,\cdots,4$; and 
\begin{align}\label{eq:gR}
g_{R{\textrm{k}}}(\zeta^2) = \lim_{a\to 0} \zeta^3 \Delta^{3/2}(\zeta^2;a) \left. \widetilde{\Gamma}_1^\ast(q^2,r^2,p^2;a) \right|_{\textrm{k}(\zeta)} \, g(a) \;,   
\end{align}
for the strong coupling, which is shown here with its explicit dependence on the subtraction scale.  

We will make, for the three-gluon vertex renormalization, the same choice made in Ref.\,\cite{Pinto-Gomez:2022brg}, that is the so-called \emph{soft-gluon} MOM scheme\footnote{Indeed, the soft-gluon kinematic configuration ($p^2=0$, $r^2=q^2$), seems to be affected by an apparent ambiguity when is defined in terms of angles, as the condition appears to be $\theta_{rp}+\theta_{qp}=\pi$, for whichever $\theta_{rp}$ and $\theta_{qp}$. Such an ambiguity can be seen not to exist as, after a careful analysis\,\cite{Aguilar:2021lke,Aguilar:2021okw}, no trace of an individual dependence on the angles remains in the limit $\theta_{pq}+\theta_{rp} \to \pi$.}; namely, $\textrm{k}(\zeta)=\textrm{s}(\zeta) \equiv \{ \zeta^2,\pi/2,\pi/2\}$. Alternatively, another natural choice for the kinematic configuration at the subtraction point might have been the symmetric one: $\textrm{k}(\zeta) = \textrm{y}(\zeta)\equiv \{ \zeta^2,2\pi/3,2\pi/3\}$; \emph{i.e.},
\begin{align}\label{eq:ZAsym}
Z_3^{\textrm{y}(\zeta)}(a) = 
\left( \widetilde{\Gamma}_1^\ast\left(\frac 2 3 \zeta^2,\frac 2 3 \zeta^2, \frac 2 3\zeta^2;a\right) \right)^{-1}  \;.    
\end{align}
However, early applications of the MOM prescription for computing the renormalized coupling \cite{Parrinello:1994wd,Alles:1996ka,Parrinello:1997wm,Boucaud:1998bq,Boucaud:1998bq}, as well as more recent analyses\,\cite{Boucaud:2013jwa,Athenodorou:2018zjv,Athenodorou:2018jwu}, have used instead a slightly different definition of the symmetric MOM scheme:  $\overline{\textrm{y}}(\zeta) \equiv \{3/2\zeta^2,2\pi/3,2\pi/3\}$; such that
\begin{align}\label{eq:ZAsymbar}
Z_3^{\overline{\textrm{y}}(\zeta)}(a) = 
\left( \widetilde{\Gamma}_1^\ast(\zeta^2,\zeta^2,\zeta^2;a)   \right)^{-1} \;. 
\end{align}
Both Eqs.\,\eqref{eq:ZAsym} and \eqref{eq:ZAsymbar} corresponding to symmetric configurations for the subtraction point, they basically differ by the choice of the invariant $s^2$, and are consequently represented by points lying on different planes in Fig.\,\ref{fig:triangle}.  

However, all renormalized form factors can be straightforwardly related to any other on the ground of Eqs.\,(\ref{eq:Z3kcdef},\ref{eq:GiR}). While, for the strong coupling, one can derive from Eq.\,\eqref{eq:gR} that 
\begin{subequations}\label{eq:grany}
\begin{align}\label{eq:granybare}
g_{R{\textrm{k}}}(\zeta^2) &= \frac{ \left. \widetilde{\Gamma}^\ast_1(q^2,r^2,p^2;a) \right|_{\textrm{k}(\zeta)}}
{\overline{\Gamma}^{\textrm{sg}}(\zeta^2;a)} \; g_{R{\textrm{s}}}(\zeta^2)  \\ 
&= \left. \widetilde{\Gamma}^\ast_{1R{\textrm{s}(\zeta)}}(q^2,r^2,p^2) \right|_{\textrm{k}(\zeta)} \; g_{R{\textrm{s}}}(\zeta^2) \,,    
\label{eq:granyren}
\end{align}    
\end{subequations}
relating three-gluon couplings in any MOM renormalization scheme $R\textrm{k}$ (with the subtraction point fixed at $k(\zeta)$) to $R\textrm{s}$ (at $s(\zeta)$; soft-gluon) through the ratio of their bare tree-level form factors, or, alternatively, the renormalized form factor which comes out from that ratio.

\section{Results}

As explained in Subsec.\,\ref{subsec:Bose}, we can apply the projector defined by Eq.\,\eqref{eq:Projdef} to extract the form factors $\widetilde{\Gamma}^\ast_i$, or equivalently its analogous associated to the tensor basis \eqref{eq:tls} for $\widetilde{\Gamma}_i$, from the transversely projected vertex calculated with lattice QCD bare Green's functions as shown in Eq.\,\eqref{eq:Gammabar}. In Ref.\,\cite{Pinto-Gomez:2022brg,Pinto-Gomez:2023zvj}, we have already presented a first careful analysis of the three-gluon form factors $\overline{\Gamma}_i$ in bisectoral kinematics from lattice QCD. A main outcome of that analysis appeared to be the emergence of a property for these form factors, very specially for the tree-level one, named therein \emph{planar degeneracy}, that can be simply described as their keeping dependence, in very good approximation, only on the $s$-invariant, Eq.\,\eqref{eq:s2}.  

Assumed beyond the bisectoral kinematics, as suggested therein\footnote{Only an exploratory study, over a reduced number of non-bisectoral kinematics configuration, have been made so far confirming the observation of the property\,\cite{Pinto-Gomez:2022brg,Aguilar:2022thg}.} and applied next to confirm the Schwinger mechanism in Ref.\,\cite{Aguilar:2022thg}, planar degeneracy implies that 
\begin{align}\label{eq:kcfromRsg}
\left. \widetilde{\Gamma}^\ast_{1R{\textrm{s}(\zeta)}}(q^2,r^2,p^2) \right|_{\textrm{k}(\zeta)} = 1 \;,  
\end{align}
as far as evaluation at ${\textrm{k}(\zeta)}$ kinematics indicates that $s^2=\zeta^2$, for any values of $\theta_{pq}$ and $\theta_{pr}$; and the scheme $R{\textrm{s}(\zeta)}$ fixes the subtraction point at $s^2=\zeta^2$ with $\theta_{pq}=\theta_{pr}=\pi/2$. In short, required to take the value of 1 by the renormalization condition at the subtraction point, the renormalized tree-level form factor takes the same value for all angles because it only keeps dependence on $s^2$ (this result is strikingly illustrated by Fig.\,3 in Ref.\,\cite{Aguilar:2022thg} for the bisectoral case).

This result \eqref{eq:kcfromRsg}, applied to Eq.\,\eqref{eq:granyren} entails that the observance of planar degeneracy is plainly equivalent to the assertion that the three-gluon coupling can be uniquely defined, irrespectively of the chosen kinematics, when the the subtraction scale is fixed for the Bose-symmetric $s$-invariant. Although approximate, this is a remarkable result.      

\begin{table}
    \centering
    \begin{tabular}{cccc}
    \hline   \hline
         $\beta$ & $(L/a)^4$ & $a$ (fm) & confs. \\
    \hline   \hline
         5.6 & 32 & 0.236 & 2000 \\
    \hline
         5.7 & 32 & 0.182 & 1000 \\
    \hline
         5.8 & 32 & 0.144 & 2000 \\
    \hline
         6.0 & 32 & 0.096 & 2000 \\
    \hline
         6.2 & 32 & 0.070 & 2000 \\
    \hline
         6.4 & 32 & 0.054 & 1290 \\
    \hline   \hline
    \end{tabular}
    \caption{Lattice setups for the gauge field configurations used in this paper. The lattice spacings in third column have been obtained using the absolute calibration reported in \cite{Necco:2001xg} at $\beta=5.8$ and supplemented by a relative calibration based in the gluon propagator scaling following the procedure described in \cite{Boucaud:2018xup}.}
    \label{tab:latticesetups}
\end{table}

In the following, we use a large-statistics set of quenched lattice-QCD configurations for multiple $\beta$'s (described in Tab.\,\ref{tab:latticesetups}) 
and investigate further the validity of the planar degeneracy by extending previous results\,\cite{Pinto-Gomez:2022brg,Pinto-Gomez:2023zvj} to general kinematics; i.e., any set of lattice momenta $(q,r,p)$ satisfying $q+r+p=0$, corresponding to the white area in Fig.~\ref{fig:triangle}. 

  The large amount of lattice momenta for these general kinematics\footnote{For a $N^4$ lattice with $N=32$, and limiting the momenta ($p=\frac{2\pi}{N a} n$) to $n\leq N/4$ to minimise the impact of lattice artifacts, there are 74800 sets of momenta for the general kinematics, 1400 for the bisectoral, and only 32 and 64 for the symmetric and soft-gluon respectively.} will allow us (i) the scrutiny of possible effects due to lattice discretization artifacts associated to the breaking of the rotational invariance\,\cite{Becirevic:1999uc,Becirevic:1999hj,deSoto:2022scb}; (ii) those beyond the planar degeneracy, i.e., residual dependencies either in $t$,$u$ in Eqs.~(\ref{eq:t4},\ref{eq:u6}) or in the angles $\theta_{qr}$, $\theta_{qp}$; and finally (iii) capitalize on the information for all the orbits to gain some physical insights into the three-gluon vertex running in its non-perturbative regime.

\subsection{H4 dependence}

It is well known that any given two-point lattice-regularized Green function, \emph{e.g.} the gluon propagator, owing to the breaking of rotational $O(4)$ down to permutation $H(4)$ group invariance, depends on the four $H(4)$-invariants\,\cite{Becirevic:1999uc,Becirevic:1999hj,deSoto:2007ht,deSoto:2022scb} 
\begin{align}
q^{[2n]}=\sum_{\mu=1}^4 q_\mu^{2n}
\end{align}
with $n=1,\dots,4$; although only the dependence on $q^{[2]}\equiv q^2$ survives in the continuum limit: Clearly, $q^{[2n]}/q^2 \sim a^{2(n-1)} \to 0$. Thus, the dominant subleading contribution when the continuum limit is approached comes with $q^{[4]}$. A procedure, usually dubbed $H(4)$-extrapolation\,\cite{Becirevic:1999uc,Becirevic:1999hj,deSoto:2007ht,deSoto:2022scb}, has been developed and proven to be notably efficacious to cure two-points Green's functions from the $O(4)$-breaking lattice artifacts\,\cite{Boucaud:2008gn,Blossier:2010ky,Blossier:2012ef,Binosi:2016nme,Boucaud:2018xup,Zafeiropoulos:2019flq,Cui:2019dwv,Pinto-Gomez:2022brg,Aguilar:2022thg}.   

In three-point cases, as the three-gluon vertex, a lattice-regularized Green's function could generally depend on any $H(4)$-invariants that can be built with three momenta. The choice of a Bose-symmetric tensor basis implies for the corresponding form factors their depending only on Bose-symmetric combinations of $H(4)$-invariants, the dominant subleading contribution coming with 
\begin{subequations}
\begin{align}\label{eq:s4}
s^{[4]} &=  \frac{q^{[4]}+r^{[4]}+p^{[4]}}{4} \\ 
&= \frac 1 2 \sum_{\mu=1}^4 q_\mu^{2}r_\mu^{2}+q_\mu^{2}p_\mu^{2}+p_\mu^{2}r_\mu^{2} \;.  
\label{eq:s4p}
\end{align}    
\end{subequations}
Other $H(4)$-invariants of $a^4$-dimensions can be also expressed in terms of $s^{[4]}$, as is made apparent by \eqref{eq:s4p}.

Thus far, the $H(4)$-extrapolation has been only successfully applied for the three-point Green's functions in the soft-gluon case, in which only one momentum scale survives and, concerning the kinematic analysis, it behaves as a two-point case\,\cite{Athenodorou:2016oyh,Boucaud:2017obn,Aguilar:2019uob,Aguilar:2021lke,Pinto-Gomez:2022brg,Aguilar:2022thg}. While, in the symmetric three-gluon case, the number of exploited kinematic configurations made unfeasible any detailed analysis of $O(4)$-breaking artifacts.  

\begin{figure}
\includegraphics[width=1.1\columnwidth]{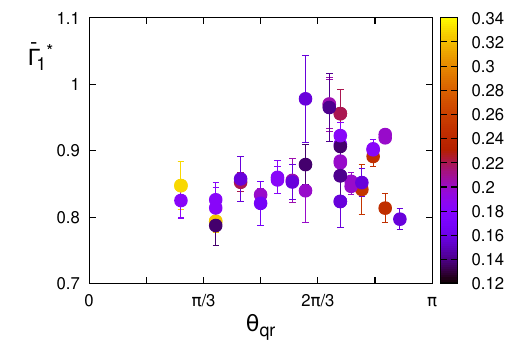} 
\caption{\label{fig:H4errors} The tree-level form factor for bisectoral kinematics ($q^2=r^2$) displayed in terms of the angle $\theta_{qr}$, for a fixed value of the Bose-symmetric invariant $s^2=84 a^{-2}(\beta)$, obtained for the lattice set-up with $\beta=6.0$ (see Tab.\,\ref{tab:latticesetups}). The color scale refers to the dimensionless ratio $s^{[4]}/(s^2)^2$.}
\end{figure}

In this work, we capitalize on the statistical power for general kinematics and perform a first analysis of the impact of these artifacts on the three-gluon vertex. The main outcome of the analysis is illustrated by Fig.~\ref{fig:H4errors}, which delivers the tree-level form factor $\overline{\Gamma}^\ast_1(q^2,p^2)$ for the bisectoral case in terms of $\theta_{qr}=2(\pi-\theta_{qp})$ and for a given $s^2$; while the value of the $H(4)$-invariant $s^{[4]}$ appears indicated by a color scale. Namely, no sizeable and systematic effect of $s^{[4]}$ appears on the tree-level form factor, as all data for almost each angle are in practice compatible within the errors. Otherwise expressed, a linear slope in $s^{[4]}$ computed for almost each angle is compatible with zero. In the case shown in the figure, the average\,\footnote{Planar degeneracy makes the average significant as it entails that the slope in $s^{[4]}$ is then  independent of the angle.} of the slopes for all the angles is -0.03(16). It should be also highlighted that both the results' errors and dispersion notably increase near the symmetric configuration, $\theta_{qr}=2\pi/3$. These observation is explained by the fact that, projecting out the form factor, one needs to deal with the inversion of a matrix, Eq.\,\eqref{eq:Mm1}, whose determinant approaches zero near the symmetric configuration, due to dimensional reduction\,\cite{Pinto-Gomez:2022brg}.   

These findings are systematically observed for the different $\beta$'s, any value of $s^2$ and all the kinematic configurations. Therefore, with the available statistical errors (corresponding to 2000 gauge field configurations), it can be concluded that the lattice $O(4)$-breaking artifacts are not relevant and,  consequently, one can reliably average all the data that only differ in $s^{[4]}$ or other higher-order $H(4)$-invariants.

\subsection{Planar degeneracy}

\begin{figure}
\centering
\includegraphics[width=\columnwidth]{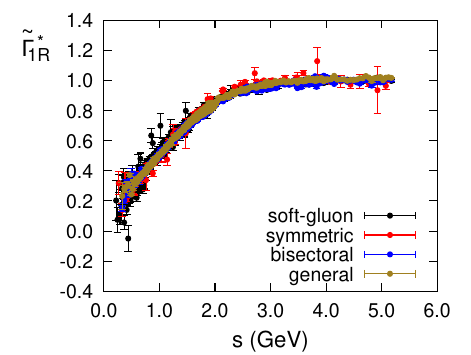}
\caption{\label{fig:Gamma1general} The tree-level form-factor evaluated at any kinematic configuration $(q^2,r^2,p^2)$ and renormalized at the soft-gluon point $s(\zeta)$, Eq.\,\eqref{eq:planar2}'s lhs, obtained from all the lattice configurations in Table \ref{tab:latticesetups}, the $O(4)$-breaking artefacts treated as explained in the text. We have taken $\zeta$=4.3 GeV. The results for all available (non-)bisectoral configurations evaluated at the same value of $s^2$ have been averaged and displayed with solid (golden) blue circles. The symmetric and soft-gluon cases are particularly shown in red and black, respectively. 
}
\end{figure}

The planar degeneracy exhibited by the three-gluon tree-level form factor in a Bose-symmetric tensor basis can be mathematically expressed as
\begin{align}\label{eq:planar}
\widetilde{\Gamma}_1^\ast(q^2,r^2,p^2;a) 
= \overline{\Gamma}^\textrm{sg}(s^2;a) \;,
\end{align}
in terms of bare quantities; which can be next renormalized in $R\textrm{s}$, the soft-gluon MOM scheme, to recast \eqref{eq:planar} as
\begin{align}\label{eq:planar2}
\widetilde{\Gamma}_{1 R\textrm{s}(\zeta)}^\ast(q^2,r^2,p^2) 
= \overline{\Gamma}_{R}^\textrm{sg}(s^2) = \frac{\overline{\Gamma}^\textrm{sg}(s^2;a)}{\overline{\Gamma}^\textrm{sg}(\zeta^2;a)} \;.
\end{align}
This is what Fig.\,\ref{fig:Gamma1general} shows from our lattice QCD calculation with the Landau-gauge field configurations described in Tab.\,\ref{tab:latticesetups}. Therein, we average and display the results for Eq.\,\eqref{eq:planar2}'s lhs (here, and in what follows, we take $\zeta$=4.3 GeV) from all the available general non-bisectoral configurations sharing the same $s$-invariant (golden filled circles); \emph{i.e.}, those evaluated at $\textrm{k}(s)$ with any available $\theta_{qp} \neq \theta_{rp}$. The tiny errors, evaluated as the mean squared width for the distribution of values at a given $s$, stem from their almost negligible dispersion and prove that Eq.\,\eqref{eq:planar2}'s lhs does depend only on $s^2$ and not on the angles. On top of this, the outputs for bisectoral (blue: $\textrm{k}(s)$ with any available $\theta_{qp}=\theta_{rp}$), symmetric (red: $\textrm{k}(s)$ with $\theta_{qp}=\theta_{rp}=2\pi/3$) and soft-gluon (black: $\textrm{k}(s)$ with $\theta_{qp}=\theta_{rp}=\pi/2$) are also displayed. The remarkable coincidence of all the results proves the equality expressed by Eq.\,\eqref{eq:planar2} and, hence, planar degeneracy within the exposed kinematic range up to $\sim$ 4-5 GeV.      

\begin{figure*}[ht]
\centering
\begin{tabular}{cc}
\includegraphics[width=\columnwidth]{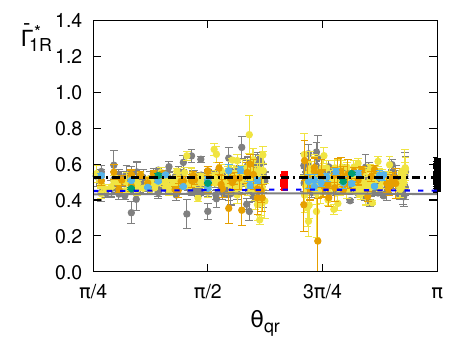}
&
\includegraphics[width=\columnwidth]{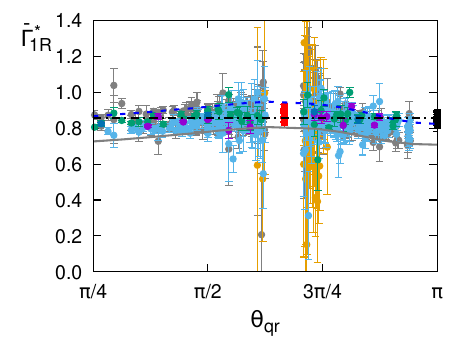}
\\
\includegraphics[width=\columnwidth]{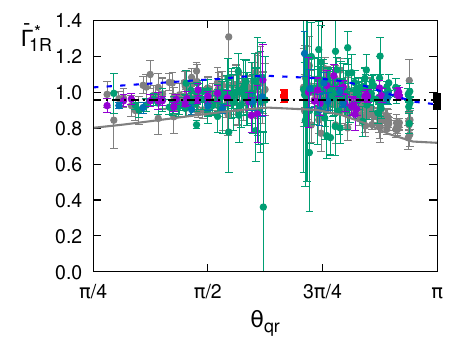}
&
\includegraphics[width=\columnwidth]{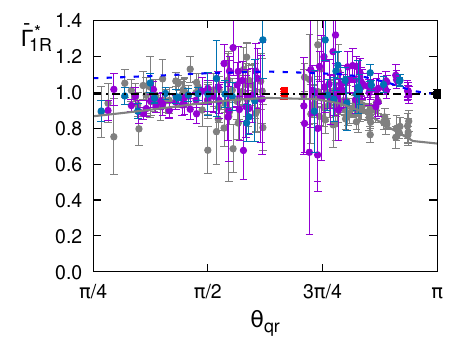}

\end{tabular}
\caption{\label{fig:Gamma1angular} The tree-level form factor $\overline{\Gamma}_1^*(q^2,p^2)$ for bisectoral kinematics ($q^2=r^2$) renormalized as in Eq.\eqref{eq:GiR}, at the soft-gluon point $s(\zeta)$ with $\zeta=4.3$ GeV, as a function of the angle between equal momenta, $\theta_{qr}$, for fixed values of $s$: from left to right $1$ and $2$ GeV (top) and $3$ and $4$ GeV (bottom). The results obtained from lattice configurations with different $\beta$'s are displayed in different colors: $\beta=5.6$ (yellow), $5.7$ (orange), $5.8$ (light blue), $6.0$ (green), $6.2$ (purple) and $6.4$ (dark blue). The same for $\overline{\Gamma}_1(q^2,p^2)$ is shown in grey, thus to expose the advantage of using the tensor basis \eqref{eq:redef} instead of \eqref{eq:tls}, basically at large $s$ and $\theta_{qr}$ near $\pi$ (soft-gluon). Results for values of $\theta_{qr}$ near $2\pi/3$ (symmetric) have been eliminated for clarity, as they are affected by the numerical noise induced by that one of matrix\,\eqref{eq:Mm1}'s eigenvalues approaches zero. The soft-gluon (symmetric) case is plotted in black (red) for all $\beta$'s. The average of soft-gluon results is also displayed by a dot-dashed black line. The DSE results obtained as discussed in Ref.\,\cite{Aguilar:2023qqd} appear displayed in each panel, at its corresponding value of $s$ with a dashed blue (solid grey) line for $\overline{\Gamma}_1^*$ ($\overline{\Gamma}_1$).}
\end{figure*}

Then, we focus on the bisectoral case, considering four bins of momenta defined by a narrow interval around $s=1$, $2$, $3$ and $4$ GeV, respectively, and display them in terms of the angle $\theta_{qr}=2(\pi - \theta_{rp})$ in the four panels of Fig.\,\ref{fig:Gamma1angular}. Eq.\,\eqref{eq:planar} establishes that, for each bin, the outputs should take the same value, which is referred to the soft-gluon case. This is what the four panels show, where the special soft-gluon and symmetric cases appear differentiated in black and red, respectively. 

Planar degeneracy implies equivalent features for the form factors in any Bose-symmetric tensor basis; particularly for those associated to the basis \eqref{eq:tls}, $\widetilde{\Gamma}_1$, which relates to $\widetilde{\Gamma}_1^\ast$ as expressed in Eq.\,\eqref{eq:GastvsG}. 
The results for $\overline{\Gamma}_1$ and $\overline{\Gamma}_1^\ast$, which derive from $\widetilde{\Gamma}_1$ and $\widetilde{\Gamma}_1^\ast$ in the bisectoral case, are shown in Fig.\,\ref{fig:Gamma1angular} (the former's in gray), and their comparison made apparent that, when the soft-gluon limit is approached and $s^2$ increases, 
the latter respects better the planar degeneracy than the former. The declining of $\overline{\Gamma}_1$ when $\theta_{qr} \to \pi$ is understood as caused by a perturbative singularity in this limit\,\cite{Aguilar:2022thg}, which can be nonperturbatively removed\,\cite{Aguilar:2023qqd} but spoils increasingly the constant behavior imposed by Eq.\,\eqref{eq:planar2} when $s^2$ grows up. This perturbative singularity also enters in $\overline{\Gamma}_3$ cancelling that of $\overline{\Gamma}_1$ when they are combined\footnote{They are indeed combined by an equation which results from applying Eq.\eqref{eq:barfromtilde} to \eqref{eq:GastvsG} and which, as $f_1^\ast=f_1+3/2 f_3$, appears to be the same restricted to $d=3$.} as in Eq.\,\eqref{eq:GastvsG}, thereby resulting in $\overline{\Gamma}_1^\ast$ which delivers by itself the correct soft-gluon limit (see Eq.\,\eqref{eq:sg}).   

All these results, obtained from lattice gauge field configurations, are plainly consistent with the analysis based on a continuum DSE calculation of the three-gluon vertex in Ref.\,\cite{Aguilar:2023qqd}. As an illustration of this excellent agreement, we have included in every panel of Fig.\,\ref{fig:Gamma1angular} the DSE result for $\overline{\Gamma}_1^\ast$ and $\overline{\Gamma}_1$ obtained at the corresponding value of $s^2$ as explained in Ref.\,\cite{Aguilar:2023qqd}.   

\begin{figure}[ht]
\centering
\begin{tabular}{c}
\includegraphics[width=\columnwidth]{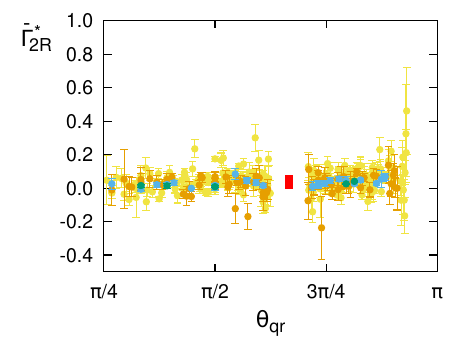}
\\
\includegraphics[width=\columnwidth]{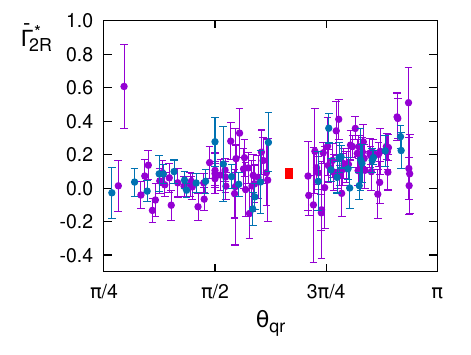}

\end{tabular}
\caption{\label{fig:Gamma2angular} The form factor $\overline{\Gamma}_2^*(q^2,p^2)$ for bisectoral kinematics ($q^2=r^2$), renormalized as in Eq.\,\eqref{eq:GiR} at the soft-gluon point $\textrm{s}(\zeta)$ with $\zeta$=4.3 GeV, for the two fixed values $s=1$ and $4$ GeV (legends as in Fig.~\ref{fig:Gamma1angular}). The results for the symmetric point are shown in red and, as in Fig.~\ref{fig:Gamma1angular}, results for $\theta_{qr}$ near $2\pi/3$ are removed to avoid numerical noise. The soft-gluon result ($\theta_{qr}=\pi$) does not exist as the corresponding basis tensor $\overline{\lambda}^*_{2\alpha\mu\nu}$ vanishes in this limit; while, in the symmetric case ($\theta_{qr}=2\pi/3$), the output is a combination of the limits for $\overline{\Gamma}_2^*(q^2,p^2)$ and $\overline{\Gamma}_3^*(q^2,p^2)$ (\emph{viz.} Eqs.\,\eqref{eq:Gamma12sym}). The form factor $\overline{\Gamma}_2(q^2,p^2)$, associated to the basis \eqref{eq:tls}, is identical to $\overline{\Gamma}_2^*(q^2,p^2)$, according to \eqref{eq:GastvsG}. 
} 
\end{figure}

\begin{figure}[ht]
\centering
\begin{tabular}{c}
\includegraphics[width=\columnwidth]{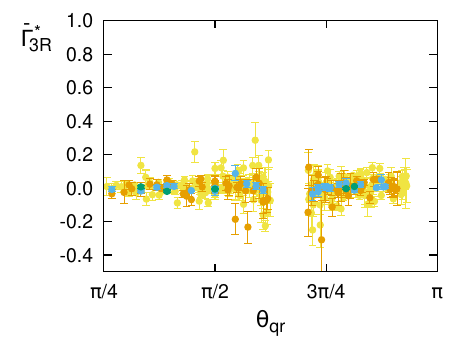} 
\\
\includegraphics[width=\columnwidth]{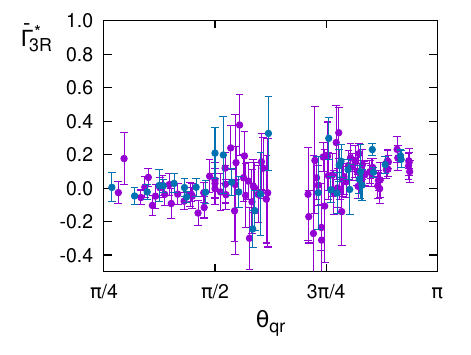}
\end{tabular}
\caption{\label{fig:Gamma3angular} The same of Fig.\,\ref{fig:Gamma2angular}, here for $\overline{\Gamma}_3^*(q^2,p^2)$. In this case, neither the soft-gluon nor the symmetric cases exist because, according to Eqs.\,\eqref{eq:Gamma12sym}, the basis tensor $\overline{\lambda}^\ast_{3\alpha\mu\nu}$ projects into the soft-gluon limit of $\overline{\lambda}^\ast_{1\alpha\mu\nu}$ or into the symmetric limits of $\overline{\lambda}^\ast_{1\alpha\mu\nu}$ and $\overline{\lambda}^\ast_{2\alpha\mu\nu}$.
}
\end{figure}

Concerning the other two form factors, $\overline{\Gamma}^\ast_2$ and $\overline{\Gamma}^\ast_3$, again derived from $\widetilde{\Gamma}^\ast_2$ and $\widetilde{\Gamma}^\ast_2$ as given by Eqs.\,\eqref{eq:barfromtilde} and, according to Eq.\,\eqref{eq:GastvsG}, equivalent to $\overline{\Gamma}_2$ and $\overline{\Gamma}_3$; we found them almost compatible with zero, within errors, and negligible thereby, as claimed in Ref.\,\cite{Pinto-Gomez:2022brg}. They are plotted in Figs.\,\ref{fig:Gamma2angular} and \ref{fig:Gamma3angular}, respectively, for $s=1$ and $4$ GeV. It can be therein noticed that the only sizeable non-zero contribution, but very small compared to $\overline{\Gamma}_1^\ast$, takes place for $\overline{\Gamma}_3^\ast$ in the soft-gluon limit ($\theta_{qr}=\pi$) when $s$ becomes large. Indeed, this contribution to $\overline{\Gamma}_3^\ast \equiv \overline{\Gamma}_3$ explains why $\overline{\Gamma}^\ast_1$ and $\overline{\Gamma}_1$ differ from each other in that kinematic domain, cancelling out the effect of the perturbative singularity in $\overline{\Gamma}_1$ and restoring the planar degeneracy as an excellent approximation for $\overline{\Gamma}_1^\ast$. 

Although choosing to display the bisectoral case, in which they depend only on a single angle making thus easier the representation, we have checked that $\overline{\Gamma}^\ast_2$ and $\overline{\Gamma}^\ast_3$ are negligible beyond, in any general kinematic configuration. Clearly, this implies the same also for $\widetilde{\Gamma}^\ast_2$, $\widetilde{\Gamma}^\ast_3$ and $\widetilde{\Gamma}^\ast_4$ or, otherwise, according to Eq.\,\eqref{eq:barfromtilde}, a very delicate and implausible cancellation needs to involve the latter with the two former ones. Furthermore, we have calculated $\widetilde{\Gamma}^\ast_4$ and concluded that, albeit noisier than the other form factors, is compatible with zero in general kinematics. 

Altogether, our findings support that one can very accurately approximate the transversely projected three-gluon vertex by
\begin{align}\label{eq:3gluonR}
\overline{\Gamma}_{R \alpha\mu\nu}(q,r,p) = \overline{\Gamma}_R^\textrm{sg}(s^2) \,\widetilde{\lambda}_{1 \alpha\mu\nu}(q,r,p) \;;   
\end{align}
renormalized at a subtraction point defined only by $s^2=\zeta^2$, irrespectively of the angles or, alternatively, the two other Bose-symmetric invariants.

\subsection{Zero crossing and logarithmic singularity}

The key ingredient defining the transversely projected three-gluon vertex, according to Eq.\,\eqref{eq:3gluonR}, is the soft-gluon form factor $\overline{\Gamma}^\textrm{sg}$. Thus far, we have carefully scrutinized the tree-level form factor, $\overline{\Gamma}_1^\ast$, and proved that planar degeneracy works remarkably well to describe its behaviour for any general kinematic configuration within the exposed range. At this point and in the following, we assume planar degeneracy and recalculate $\overline{\Gamma}^\textrm{sg}$ from lattice QCD by capitalizing on our large-statistics sample of gauge configurations, as an average over the many different kinematic configurations $k(s)$, for all available angles $\theta_{qp}$ and $\theta_{qr}$ at each value of the Bose-symmetric invariant $s$. 

\begin{figure}[ht]
\centering
\includegraphics[width=\columnwidth]{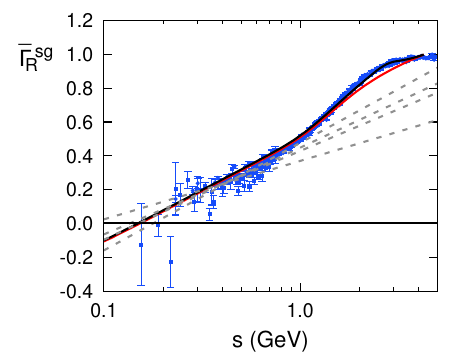} 
\caption{\label{fig:Gamma1logslope} Semi-log plot of the form factor $\overline{\Gamma}_R^\textrm{sg}$ evaluated, as explained in the text, by assuming planar degeneracy and averaging for all kinematic configurations at fixed $s^2$. 
The fit given in Ref.\,\cite{Aguilar:2021okw} (\emph{viz.} Eq.\,\eqref{eq:fitR}) is displayed in solid red line, while the refined version herein introduced (\emph{viz.} Eq.\,\eqref{eq:fitG}) is in black; and the fits with Eq.\,\eqref{eq:IRfit} described in Tab.\,\ref{tab:slopes} are depicted with dashed grey lines.
}
\end{figure}

Fig.\,\ref{fig:Gamma1logslope} shows the lattice results for $\overline{\Gamma}_R^\textrm{sg}$ and displays also a best-fit with an expression motivated by a construction of the three-gluon vertex relying on the STIs that it satisfies\,\cite{Aguilar:2019jsj,Aguilar:2019kxz,Aguilar:2021lke,Aguilar:2021okw}, Eqs.\,(\ref{eq:fitR},\ref{eq:fitG}) in appendix\,\ref{app:fit}. A main feature of the three-gluon vertex revealed by this construction, and confirmed by lattice analyses~\cite{Athenodorou:2016oyh,Duarte:2016ieu,Boucaud:2017obn}, is the presence of an infrared zero-crossing at very low values of $s$ induced by a logarithmic divergence at vanishing momenta. This singularity plays an important dynamical role as it is caused by the interplay of massless ghosts and massive gluons in the loops of the three-gluon DSE expansion\,\cite{Aguilar:2013vaa}, and can be related to the dynamical gluon mass generation\,\cite{Binosi:2012sj,Aguilar:2011xe} at the level of the two-point function \emph{via} the corresponding STI\,\cite{Aguilar:2019jsj} (see appendix\,\ref{app:fit}). 

\begin{table}[ht]
    \begin{tabular}{|c|c|c|}
    \hline
    \hline
        $s_\text{max}$ (GeV) & $\gamma_1$ & $s_0$ (MeV) \\
    \hline
    \hline
         0.7 & 0.138(2) & 178(3) \\
    \hline
         0.6 & 0.118(6) & 150(9) \\
    \hline
         0.5 & 0.107(16) & 135(23)\\
    \hline
         0.4 & 0.075(23) & 85(38)\\
    \hline
    \hline
    - & 0.11 & 161 \\
    \hline
    \end{tabular}
    \caption{Logarithmic slope $\gamma_1$ and zero crossing location $s_0$ for different fitting windows and combining data for all the kinematics and lattice setups. The last row corresponds to the results from the best-fit of Eq.\,\eqref{eq:fitG} to all data.}
    \label{tab:slopes}
\end{table}

Our best-fit over the ensemble of lattice data leaves us with a global  determination for the zero-crossing location, as shown in Tab.\,\ref{tab:slopes}. However, the fitting expression behaves as    
\begin{equation}\label{eq:IRfit}
    \overline{\Gamma}_{R}^\textrm{sg}(s^2) = \gamma_0 + \gamma_1 \ln\left(\frac{s^2}{\zeta^2}\right)\,,
\end{equation}
as asymptotically low $s$, with $\gamma_1$=0.11 (\emph{viz.} Eq.\,\eqref{eq:figGas}). Indeed, Eq.\,\eqref{eq:IRfit} is the more general output from identifying and linking \emph{via} STI the two- and three-point singularities generated by massless ghosts\,\cite{Aguilar:2011xe,Athenodorou:2016gsa}. The same can be also concluded from other approaches for the description of the low-momentum behavior of QCD Green's functions\,\cite{Pelaez:2014mxa}. Therefore, in the aim of estimating a systematic uncertainty for the zero-crossing location, we fit Eq.\,\eqref{eq:IRfit} with free parameters to all the lattice data below a given scale $s_\text{max}$, ranging from $0.4$ to $0.7$ GeV, and derive thereby the location from the fitted parameters: $s_0=\zeta \exp{(-\gamma_0/[2\gamma_1])}$. The results can be found in Tab.\,\ref{tab:slopes} and the corresponding fits appear displayed in Fig.\,\ref{fig:Gamma1logslope}. It should be noticed that the slope gets smaller for smaller fitting windows, likely indicating that the uncertainty for a lower zero-crossing location values is underestimated.  

\begin{figure}[ht]
\centering
\includegraphics[width=\columnwidth]{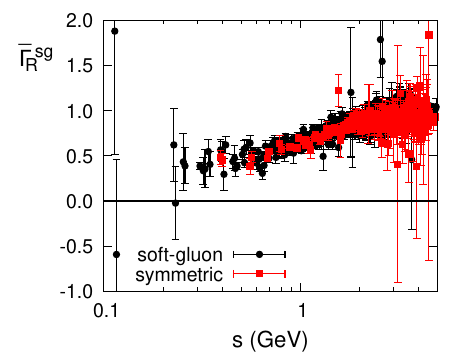}
\caption{\label{fig:Gamma1DWF} The tree-level form factor renormalized at the soft-gluon point $\textrm{s}(\zeta)$ with $\zeta$=4.3 GeV, Eq.\,\eqref{eq:planar2}'s lhs, obtained from $\overline{\Gamma}^\textrm{sym}_1$ and $\overline{\Gamma}^\textrm{sg}$, see Eqs.\,\eqref{eq:Gamma12sym} and \eqref{eq:Gammasg}, calculated with unquenched $N_F=2+1$ lattice configurations at a physical pion mass\,\cite{Aguilar:2019uob}, and here expressed in terms of the Bose-symmetric invariant $s^2$.}
\end{figure}

The tree-level form factor has been formerly analyzed in both the particular cases of symmetric and soft-gluon kinematics\,\cite{Aguilar:2019uob}, by using a set of $N_F=2+1$ DWF unquenched gauge-field lattice configurations at a physical pion mass, to unveil any possible effect from realistic dynamical quarks. Then, we capitalize on the results of Ref.\,\cite{Aguilar:2019uob} to test and confirm that planar degeneracy works equally well in the unquenched case, by displaying them in terms of the Bose-symmetric invariant $s$ in Fig.\,\ref{fig:Gamma1DWF}: lattice data for the tree-level form factor in the two different kinematic cases, both renormalized at the same subtraction point, exhibit an almost perfect overlapping within errors. As for the quenched case, the agreement is indeed perfect within the IR domain and up to roughly 4 GeV, some marginal discrepancies appearing above which should rely on the distinct perturbative running for different kinematics. 

\begin{figure}[t]   
\centering
\includegraphics[width=\columnwidth]{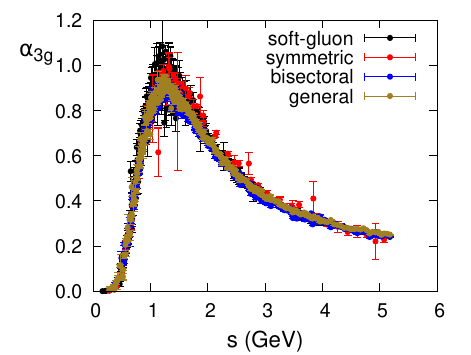}
\caption{\label{fig:alpha} Strong effective coupling defined by Eq.\,\eqref{eq:alpha3g} and obtained according to
Eq.(\ref{eq:grany}) for all kinematics as done (and labelled) in Fig.\,\ref{fig:Gamma1general}.
 }
\end{figure}

Finally, the four panels of Fig.\,\eqref{fig:Gamma1angular} made strikingly apparent the validity of Eq.\,\eqref{eq:planar} from our lattice calculation, and hence \eqref{eq:kcfromRsg}. The DSE results therein displayed also confirms that this is a good approximation and, specially within the IR domain, becomes exact in practice (see the first panel for $s$=1 GeV). We can therefore conclude that, in very good approximation, a unique three-gluon coupling
\begin{align}\label{eq:alpha3g}
\alpha_{3g}(\zeta^2) = \frac{g^2_{R\textrm{k}}(\zeta^2)}{4\pi} \equiv \frac{g^2_{R\textrm{s}}(\zeta^2)}{4\pi}    
\end{align}
results from \eqref{eq:grany}, the subtraction point fixed at $s^2=\zeta^2$ for any kinematic configuration. This is shown in Fig.\,\ref{fig:alpha}, where our lattice data display this unique three-gluon coupling behavior for momenta below $\sim$ 4 GeV. At this point, we can recall the instructive discussion in the end of section IV of Ref.\,\cite{Aguilar:2021okw}, where the effective strength associated to gluon-gluon interaction was compared to ghost-gluon by using the coupling \eqref{eq:alpha3g} in the soft-gluon renormalization scheme. Our result here, inferred from the observance of the three-gluon planar degeneracy, entails that this is the unique effective strength for the gluon-gluon interaction from the three-gluon coupling.

\section{Conclusions}

In the present investigation of the Landau-gauge three-gluon vertex from lattice QCD, we have aimed at the completion of a series of previous works by the study of the full kinematic range and tensor space available to the transversely projected vertex. 

Two different Bose-symmetric basis have been considered, both including the tree-level (classical) tensor as one of their elements. The first basis is the one implemented in the previous lattice study\,\cite{Pinto-Gomez:2022brg}, while the second one\,\cite{Aguilar:2023qqd} is built by rearranging the tensors of the first basis, only entailing a minimal modification of the form factor associated to the classical one. This classical form factor becomes then improved as it fully contains the three-gluon vertex in the soft-gluon limit. In both cases, the corresponding form factors have been projected out from the three-gluon Green's function and, evaluating carefully the impact of $O(4)$-breaking lattice artifacts, have been seen to (i) be dominated by the classical form factor; which (ii) is furthermore shown to depend only on the Bose-symmetric combination of momenta $s^2$ [\emph{viz.} Eq.\,\eqref{eq:s2}], for $s \leq$ 4-5 GeV. The latter being true in practice after the implementation of the two basis, a very small violation of (ii) in the vicinity of the soft-gluon limit and growing up with the momentum appears to be eliminated by the improved version of the basis. These results deliver an important confirmation, for a fully general three-gluon vertex, of the property unveiled in previous works and termed planar degeneracy. We have furthermore presented the first evidences for this phenomenon in realistic QCD, beyond pure Yang-Mills, by exploiting unquenched gauge field configurations using DWF at a physical pion mass. 

Finally, we have deepened into the implications of this remarkable property, and have shown its implying that the transversely projected three-gluon vertex can be uniquely renormalized in MOM scheme, irrespectively from any chosen kinematic configuration, when the subtraction point is defined by imposing a given scale for $s^2$ [\emph{viz.} Eq.\,\eqref{eq:3gluonR}]. Thus, this renormalized vertex can be expressed in terms of the only form factor in the soft-gluon case which, capitalizing on the planar degeneracy, can be very accurately displayed by considering all the lattice data for all kinematic configurations [\emph{viz.} Fig.\,\ref{fig:Gamma1logslope}]. Finally, an effective coupling can be also defined [\emph{viz.} Eq.\,\eqref{eq:alpha3g}], relying on this unique MOM three-gluon vertex, as a sensible expression of the gluon-gluon interaction strength.


\section*{Acknowledgement}

We are grateful to A.C. Aguilar, M.N. Ferreira and J. Papavassiliou, for the fruitful discussions that gave rise to some of the ideas exposed in this work; and for providing us with the DSE results displayed for comparison in Fig.\,\ref{fig:Gamma1angular}. F.P.G also thanks A.C. Aguilar for her warm hospitality and support, and Banco Santander for partially financing this work. The work has been funded by the Spanish ``Ministerio de Ciencia e Innovaci\'on (MICINN)'' through grants PID2019-107844-GB-C22 and PID2022-140440-NB-C22. The authors acknowledge the C3UPO of the Pablo de Olavide University for the support with HPC facilities.

\appendix

\section{The three-gluon STIs and the tree-level form factor}
\label{app:fit}

As discussed in, \emph{e.g.}, Ref.\,\cite{Aguilar:2021lke}, one can generally write the 1PI three-gluon vertex as
\begin{align}\label{eq:fatG}
\fatg^{\alpha\mu\nu} (q,r,p) =& V^{\alpha\mu\nu} (q,r,p) + \Gamma^{\alpha\mu\nu} (q,r,p) \;, \\
\Gamma^{\alpha\mu\nu} (q,r,p) =& \sum_{i=1}^{10} X_i(q^2,r^2,p^2) \, \ell_i^{\alpha\mu\nu}  \nonumber \\
&+ \sum_{j=1}^{4} Y_j(q^2,r^2,p^2) \, t_j^{\alpha\mu\nu} 
\label{eq:XandY}
\end{align}
where $V^{\alpha\mu\nu}$ contains the longitudinally-coupled Schwinger poles while, resorting to the Ball-Chiu basis\,\cite{Ball:1980ax,Ball:1980ay} (see, \emph{e.g.}, Eqs.\,(3.4) and (3.6) of Ref.\,\cite{Aguilar:2019jsj}), the free-pole part $\Gamma^{\alpha\mu\nu}$ is decomposed into two, transverse and non-transverse pieces. One can then capitalize on the STIs involving the three-gluon vertex, namely 
\begin{align}\label{eq:STI}
p_\nu \fatg^{\alpha\mu\nu}(q,r,p) = F(p^2) \left[ {\cal T}^{\mu\alpha}(r,p,q) - {\cal T}^{\alpha\mu}(q,p,r) \right] \;,    
\end{align}
with $F(p^2)$ standing for the ghost dressing function and ${\cal T}^{\alpha\mu}$ for a tensor structure involving the gluon propagator and the ghost-gluon scattering kernel\,\cite{Ball:1980ax,Davydychev:1996pb,Aguilar:2018csq}; and rely on Eq.\,\eqref{eq:STI}'s rhs for the construction of the non-transverse components resulting from the vertex contraction in \eqref{eq:STI}'s lhs. 

Rooting on the Schwinger mechanism for the generation of a dynamical gluon mass, the gluon propagator can be written as\,\footnote{The dependence on the renormalization scale $\zeta^2$ is kept implicit.}
\begin{align}
\left[\Delta(p^2)\right]^{-1} = p^2 J(p^2) + m^2(p^2) \;; \end{align}
and it can be proven that, in \eqref{eq:STI}, the dynamical gluon mass $m^2(p^2)$ is only attached to $V^{\alpha\mu\nu}$, and the kinetic term $J(p^2)$ to the non-transverse piece of $\Gamma^{\alpha\mu\nu}$\,\cite{Binosi:2012sj,Aguilar:2011xe}. Notwithstanding this decoupling of $m^2(p^2)$ and $J(p^2)$ at the level of the STIs, they remain coupled by the gluon propagator DSE, in which the full three-gluon vertex enters and triggers the Schwinger mechanism. 

The form factors $X_i$ defined in Eq.\,\eqref{eq:XandY} can be thus derived from the STIs like \eqref{eq:STI}, being then particularly related to $F(p^2)$ and $J(p^2)$. This latter can be seen to receive the contribution of a logarithmic singularity caused by ghost loops in the SDE expansion of the gluon propagator which, contrarily to gluon loops, do not remain \emph{protected} by a dynamically-generated mass\,\cite{Aguilar:2013vaa}. The same happens for the form factors $X_i$, which are impacted by a logarithmic singularity of the same origin from the three-gluon vertex SDE. The STIs connect both singularities\,\cite{Aguilar:2019jsj}. 

In this work, we calculate from lattice QCD the transversely projected three-gluon vertex, Eq.\,\eqref{eq:Gammabardef}. For this, the contraction of the 1PI vertex with the transverse projectors mixes the transverse and non-transverse pieces of Eq.\,\eqref{eq:XandY}, and makes the form factors in the Bose-symmetric tensor basis, either \eqref{eq:tls} or \eqref{eq:redef}, depend on different combinations of $X_i$ and $Y_i$. However, although the form factors $Y_i$ cannot be extracted from the STIs, one is specially left with\,\cite{Aguilar:2021okw}
\begin{align}
\overline{\Gamma}^\textrm{sg}(q^2) = X_1(q^2,q^2,0) - q^2 X_3(q^2,q^2,0) \;,     
\end{align}
in the soft-gluon case. Consequently, on the basis of the planar degeneracy approximation and within its validity momentum range, the transversely projected vertex remains completely defined by the STIs. 

The knowledge of $J(p^2)$ is required, and it derives from the gluon propagator DSE which, in its turn, involves the three-gluon vertex \eqref{eq:fatG}. Given the complexity of solving the gluon propagator DSE \emph{in its full glory}, an approximate iterative procedure resorting to the dynamical equation for $m^2(p^2)$\,\cite{Aguilar:2019kxz} and to gluon propagator lattice data is described and followed in Ref.\,\cite{Aguilar:2021lke}. Relying on this output for $J(p^2)$, the STI-based construction of $\overline{\Gamma}^\textrm{sg}$ appears to be in good agreement with the lattice data therein delivered for the soft-gluon case. In Ref.\,\cite{Aguilar:2021okw}, the STI-based result for $\overline{\Gamma}^\textrm{sg}$ is accurately fitted to
\begin{align}
R(p^2) &= F(p^2) T(p^2) \nonumber \\ 
&+ \nu_1 \left(\frac{1}{1+(p^2/\nu_2)^2} - \frac{1}{1+(\zeta^2/\nu_2)^2} \right) \;;  
\label{eq:fitR}
\end{align}
$R(p^2)$ being a functional form inspired by the STI construction of the vertex, with $\nu_{1,2}$=0.165, 0.83 GeV$^2$ and where 
\begin{align}
T(p^2) = 1 &+ \frac{3\lambda_s}{4\pi} \left( 1 + \frac{\tau_{1s}}{\tau_{2s}+p^2}\right) \nonumber \\
&\times \left[ 2 \ln{\frac{p^2+\eta_s^2(p^2)}{\zeta^2+\eta_s^2(\zeta^2)}} + \frac 1 6 \ln{\frac{p^2}{\zeta^2}} \right]
\label{eq:fitT}
\end{align}
and 
\begin{align}\label{eq:fitF}
F^{-1}(p^2) = 1 &+ \frac{9\lambda_F}{16\pi} \left( 1 + \frac{\tau_{1F}}{\tau_{2F}+p^2}\right) \ln{\frac{p^2+\eta_F^2(p^2)}{\zeta^2+\eta_F^2(\zeta^2)}} 
\end{align}
dominate the low-momentum behaviour and are determined by the best fits to the solution of $J(p^2)$ and to ghost dressing lattice data, respectively; with 
\begin{align}\label{eq:fiteta}
\eta^2_{s,F}(p^2) = \frac{\eta_{1s,F}^2}{1+p^2/\eta^2_{2s,F}} \;.    
\end{align}
The parameters for Eqs.\,(\ref{eq:fitT},\ref{eq:fitF}) are\,\cite{Aguilar:2021okw}: $\lambda_s$=0.27, $\tau_{1s}$=2.67, $\tau_{2s}$=1.05, $\eta^2_{1s}$=3.10, $\eta^2_{2s}$=0.729; $\lambda_F$=0.22, $\tau_{1F}$=6.34, $\tau_{2F}$=2.85, $\eta^2_{1F}$=0.107, $\eta^2_{2F}$=11.2; all parameters are in units of GeV$^2$, except the dimensionless $\lambda_{s,F}$. 

In this work, we deliver a very precise lattice prediction of the soft-gluon form factor based on planar degeneracy, and on exploiting thereby all available kinematic configurations. Consequently, we need to slightly refine the fitting expression, now reading     
\begin{align}
\overline{\Gamma}_R^\textrm{sg}(p^2) = R(p^2) &+  \frac{\nu_3}{1+[(p^2-p_0^2)/\nu_4]^2} \nonumber \\ 
&- \frac{\nu_3}{1+[(\zeta^2-p_0^2)/\nu_4]^2} \;,
\label{eq:fitG}
\end{align}
and can thus deliver a very accurate description of our prediction 
with $p_0$=2.7 GeV and $\nu_{3,4}$=0.054, 4.12 GeV$^2$, for all momenta below the subtraction point $\zeta$=4.3 GeV. It is worthwhile to remark that the low-momentum asymptotic behavior reads
\begin{align}\label{eq:figGas}
\overline{\Gamma}_R^\textrm{sg}(p^2) \sim F(0) \frac{\lambda_s}{8\pi} \left(1+\frac{\tau_{1s}}{\tau_{2s}}\right) \ln{\frac{p^2}{\zeta^2}} = 0.11 \ln{\frac{p^2}{\zeta^2}} \;, 
\end{align}
fully relying on Eqs.\,(\ref{eq:fitT},\ref{eq:fitF}) and on the DSE determinations of the two-point gluon and ghost Green's functions from Refs.\cite{Aguilar:2021lke,Aguilar:2021okw}.

\bibliographystyle{unsrt}
\bibliography{refs.bib}

\end{document}